\documentclass[aps,prc,twocolumn,showpacs,superscriptaddress,groupaddress,amsmath,amssymb]{revtex4-1}
\usepackage{graphicx}  
\usepackage{dcolumn}   
\usepackage{bm}        
\usepackage{verbatim}   
\usepackage{textcomp}

\usepackage{xcolor} 

\begin{document}

\title{Reexamination of $^{6}$Li scattering as a Probe to Investigate the Isoscalar Giant Resonances in Nuclei}

\author{J. C.~Zamora}
  \affiliation{Instituto de Fisica, Universidade de Sao Paulo, SP 05508-090,
  Brazil}

\author{C.~Sullivan}
\affiliation{National Superconducting Cyclotron Laboratory, Michigan State University, East Lansing, MI 48824,
  USA}
\affiliation{Joint Institute for Nuclear Astrophysics: CEE,  Michigan State University, East Lansing, MI 48824,
USA}
\affiliation{Department of Physics and Astronomy, Michigan State University, East Lansing, MI 48824, USA}

\author{R.G.T.~Zegers}
\affiliation{National Superconducting Cyclotron Laboratory, Michigan State University, East Lansing, MI 48824,
  USA}
\affiliation{Joint Institute for Nuclear Astrophysics: CEE,  Michigan State University, East Lansing, MI 48824,
USA}
\affiliation{Department of Physics and Astronomy, Michigan State University, East Lansing, MI 48824, USA}

\author{N.~Aoi}
\affiliation{Research Center for Nuclear Physics (RCNP), Osaka University, Ibaraki, Osaka 567-0047, Japan}


\author{L.~Batail}
\affiliation{CEA, DAM, DIF, F-91297 Arpajon, France}

\author{D.~Bazin}
\affiliation{National Superconducting Cyclotron Laboratory, Michigan State University, East Lansing, MI 48824,
  USA}
\affiliation{Department of Physics and Astronomy, Michigan State University, East Lansing, MI 48824, USA}

\author{M.~Carpenter}
\affiliation{Argonne National Laboratory, Argonne, Illinois 60439, USA}

\author{J.J.~Carroll}
\affiliation{CCDC/Army Research Laboratory, 2800 Powder Mill Road, Adelphi, Maryland 20783, USA}

\author{I.~Deloncle}
\affiliation{IJCLab, CNRS Université Paris-Saclay, F-91405 Orsay Cedex, France}
\affiliation{CEA, DAM, DIF, F-91297 Arpajon, France}

\author{Y.D.~Fang}
\affiliation{Research Center for Nuclear Physics (RCNP), Osaka University, Ibaraki, Osaka 567-0047, Japan}

\author{H.~Fujita}
\affiliation{Research Center for Nuclear Physics (RCNP), Osaka University, Ibaraki, Osaka 567-0047, Japan}

\author{U.~Garg}
\affiliation{Department of Physics, University of Notre Dame, Notre Dame, Indiana 46556, USA}

\author{G.~Gey}
\affiliation{Research Center for Nuclear Physics (RCNP), Osaka University, Ibaraki, Osaka 567-0047, Japan}

\author{C.J.~Guess}
\altaffiliation{Present address: Department of Physics and Astronomy, Rowan University, Glassboro, NJ 08028, USA}
\affiliation{Department of Physics and Astronomy, Swarthmore College, Swarthmore, Pennsylvania 19081, USA}

\author{M.N.~Harakeh}
\affiliation{Research Center for Nuclear Physics (RCNP), Osaka University, Ibaraki, Osaka 567-0047, Japan}
\affiliation{KVI-CART, University of Groningen, 9747 AA Groningen, The Netherlands}

\author{T.H.~Hoang}
\affiliation{Research Center for Nuclear Physics (RCNP), Osaka University, Ibaraki, Osaka 567-0047, Japan}

\author{E.~Hudson}
\affiliation{Department of Physics and Astronomy, Swarthmore College, Swarthmore, Pennsylvania 19081, USA}

\author{N.~Ichige}
\affiliation{Department of Physics, Tohoku University, Sendai 980-8578, Japan}

\author{E.~Ideguchi}
\affiliation{Research Center for Nuclear Physics (RCNP), Osaka University, Ibaraki, Osaka 567-0047, Japan}

\author{A.~Inoue}
\affiliation{Research Center for Nuclear Physics (RCNP), Osaka University, Ibaraki, Osaka 567-0047, Japan}

\author{J.~Isaak}
\affiliation{Research Center for Nuclear Physics (RCNP), Osaka University, Ibaraki, Osaka 567-0047, Japan}
\affiliation{Institute for Nuclear Physics, Technical University Darmstadt, D-64289 Darmstadt, Germany}

\author{C.~Iwamoto}
\affiliation{Center for Nuclear Study, University of Tokyo (CNS) RIKEN Campus, 2-1 Hirosawa, Wako, Saitama 351-0198, Japan}

\author{C.~Kacir}
\affiliation{Department of Physics and Astronomy, Swarthmore College, Swarthmore, Pennsylvania 19081, USA}

\author{N.~Kobayashi}
\affiliation{Research Center for Nuclear Physics (RCNP), Osaka University, Ibaraki, Osaka 567-0047, Japan}

\author{T.~Koike}
\affiliation{Department of Physics, Tohoku University, Sendai 980-8578, Japan}

\author{M.~Kumar~Raju}
\affiliation{Research Center for Nuclear Physics (RCNP), Osaka University, Ibaraki, Osaka 567-0047, Japan}

\author{S.~Lipschutz}
\affiliation{National Superconducting Cyclotron Laboratory, Michigan State University, East Lansing, MI 48824,
  USA}
\affiliation{Joint Institute for Nuclear Astrophysics: CEE,  Michigan State University, East Lansing, MI 48824,
USA}
\affiliation{Department of Physics and Astronomy, Michigan State University, East Lansing, MI 48824, USA}
 
\author{M.~Liu}
\affiliation{Institute of Modern Physics, Chinese Academy of Sciences, Lanzhou, China}

\author{P.~von~Neumann-Cosel}
\affiliation{Institute for Nuclear Physics, Technical University Darmstadt, D-64289 Darmstadt, Germany}

\author{S.~Noji}
\affiliation{National Superconducting Cyclotron Laboratory, Michigan State University, East Lansing, MI 48824,
  USA}
\affiliation{Joint Institute for Nuclear Astrophysics: CEE,  Michigan State University, East Lansing, MI 48824,
USA}

\author{H.J.~Ong}
\affiliation{Research Center for Nuclear Physics (RCNP), Osaka University, Ibaraki, Osaka 567-0047, Japan}

\author{S.~P\'eru}
\affiliation{CEA, DAM, DIF, F-91297 Arpajon, France}

\author{J.~Pereira}
\affiliation{National Superconducting Cyclotron Laboratory, Michigan State University, East Lansing, MI 48824,
  USA}
\affiliation{Joint Institute for Nuclear Astrophysics: CEE,  Michigan State University, East Lansing, MI 48824,
USA}

\author{J.~Schmitt}
\affiliation{National Superconducting Cyclotron Laboratory, Michigan State University, East Lansing, MI 48824,
  USA}
\affiliation{Joint Institute for Nuclear Astrophysics: CEE,  Michigan State University, East Lansing, MI 48824,
USA}
\affiliation{Department of Physics and Astronomy, Michigan State University, East Lansing, MI 48824, USA}

\author{A.~Tamii}
\affiliation{Research Center for Nuclear Physics (RCNP), Osaka University, Ibaraki, Osaka 567-0047, Japan}

\author{R.~Titus}
\affiliation{National Superconducting Cyclotron Laboratory, Michigan State University, East Lansing, MI 48824,
  USA}
\affiliation{Joint Institute for Nuclear Astrophysics: CEE,  Michigan State University, East Lansing, MI 48824,
USA}
\affiliation{Department of Physics and Astronomy, Michigan State University, East Lansing, MI 48824, USA}

\author{V.~Werner}
\affiliation{Institute for Nuclear Physics, Technical University Darmstadt, D-64289 Darmstadt, Germany}

\author{Y.~Yamamoto}
\affiliation{Research Center for Nuclear Physics (RCNP), Osaka University, Ibaraki, Osaka 567-0047, Japan}

\author{X.~Zhou}
\affiliation{Institute of Modern Physics, Chinese Academy of Sciences, Lanzhou, China}

\author{S.~Zhu}
\affiliation{Argonne National Laboratory, Argonne, Illinois 60439, USA}


\date{\today}

\begin{abstract}
Inelastic ${}^{6}$Li scattering  at 100~MeV/u on ${}^{12}$C and ${}^{93}$Nb  have been measured with the high-resolution magnetic spectrometer Grand Raiden. The magnetic-rigidity settings of the spectrometer covered excitation energies from 10 to 40~MeV and scattering angles in the range $0^\circ < \theta_{\text{lab.}}< 2^\circ$. The isoscalar giant monopole resonance was selectively excited in the present data.  Measurements free of instrumental background  and the  very favorable resonance-to-continuum ratio of ${}^{6}$Li scattering allowed for precise determination of the $E0$ strengths in ${}^{12}$C and ${}^{93}$Nb. It was found that the monopole strength in ${}^{12}$C exhausts  $52 \pm 3^\text{(stat.)} \pm 8 ^\text{(sys.)}$\% of the energy-weighted sum rule (EWSR), which is considerably higher than results from previous $\alpha$-scattering experiments. The monopole strength in ${}^{93}$Nb exhausts  $92 \pm 4^\text{(stat.)} \pm 10 ^\text{(sys.)}$\% of the EWSR, and it is consistent with measurements of nuclei  with mass number of $A\approx90$. Such comparison indicates that the isoscalar giant monopole resonance distributions in these nuclei are very similar, and no influence due to nuclear structure was observed. 
\end{abstract}

 
\maketitle

\section{Introduction}
The incompressibility of nuclear matter $K_\text{nm}$ is a fundamental quantity and an important parameter of the nuclear equation of state (EoS) with significant consequences in theories of nucleus-nucleus collisions \cite{BERTSCH1988189},  astrophysical phenomena such as supernova explosions, and properties of dense objects like neutron stars \cite{RevModPhys.89.015007}. Measurements of the isoscalar giant monopole resonance (ISGMR)  in combination with microscopic calculations provide an effective method to constrain experimentally the nuclear-matter incompressibility \cite{BLAIZOT1980171, PhysRevLett.82.691, GARG201855}.\par

Due to the scalar-isoscalar nature of the $\alpha$-particle,  compression modes such as ISGMR are predominantly excited in inelastic $\alpha$-scattering experiments at small angles. The use of the  $\alpha$-particle probe is a well-established technique that has been extensively employed  in the investigation of  isoscalar giant resonances in a wide range of nuclei \cite{PhysRevLett.38.676,HARAKEH1979373,PhysRevC.24.884, PhysRevC.33.1116,PhysRevC.64.064308,PhysRevC.73.014314,PhysRevLett.99.162503,PATEL2012447}. Also, $d$ and ${}^{6}$Li probes have been used in the past, with similar good results, in studies of isoscalar giant resonances \cite{PATEL2014387,PhysRevC.80.014312}. In particular, ${}^{6}$Li experiments have an important advantage  because of the better ratio between the resonance peak and the continuum \cite{PhysRevC.52.3195}. As ${}^{6}$Li has a low particle emission threshold ($S_\alpha =1.47$~MeV),   the breakup probability of the projectile is enhanced  with the dominant channel $d+\alpha$.    This reduces considerably the background component from the continuum and provides a better way to extract the giant resonance strengths.  In addition to the continuum background, usually in inelastic scattering experiments at scattering angles near $0^\circ$, instrumental background is also present due to beam halo or scattering of unreacted beam near the focal plane of the spectrometer. Sometimes, the combination of the continuum and instrumental background is subtracted through a parameterization \cite{BONIN1984349,PhysRevC.64.064308,PhysRevC.73.014314, PhysRevC.80.014312}. However, this parameterization carries a significant systematic error, which is difficult to estimate and can result in a substantial uncertainty in the extracted giant resonance parameters. Alternatively, the instrumental background can be subtracted by operating the spectrometer used for detecting the inelastically scattered particles in a double-focusing mode \cite{PhysRevC.68.064602,UCHIDA200312,PhysRevC.69.051301,PATEL2012447,GUPTA2016482}. In this mode, the background events exhibit a broad (and usually flat) distribution as a function of the non-dispersive angle, while  events due to scattering from target have a pronounced distribution that peak at $0^\circ$. By using a side-band analysis, excitation-energy spectra can be extracted from which the instrumental background has been reliably removed \cite{PhysRevC.68.064602,UCHIDA200312,PhysRevC.69.051301,PATEL2012447,GUPTA2016482}.  The contributions from giant resonances associated with different units of angular-momentum transfer are then obtained by using a multipole decomposition analysis (see below). \par

In this work, the ISGMRs in ${}^{12}$C and ${}^{93}$Nb were investigated via ${}^{6}$Li scattering experiments. Although the  $E0$ strength of ${}^{12}$C has been studied in the past by using different probes (${}^{3}$He, $\alpha$, ${}^{6}$Li) \cite{PhysRevC.24.2720,PhysRevC.36.416,PhysRevC.57.2748,PhysRevC.68.014305},  the impact of the background subtraction on the extracted multipole strengths  is still  not well understood. In the ${}^{6}$Li experiment reported  here, no instrumental background was present in the measurements around $0^\circ$ scattering angle, and a subtraction through a parameterization of the instrumental background or through a side-band analysis was not necessary. In combination with the favorable ratio of resonance-to-continuum background, the $E0$ transition strengths could be reliably extracted.
Also, recent $\alpha$-scattering measurements on $A\approx90$ nuclei opened a discussion about the nuclear-structure influence on the centroid energy of the ISGMR for nuclei in this mass region \cite{PhysRevC.88.021301,GUPTA2016482}.
The ${}^{93}\text{Nb}({}^{6}\text{Li},{}^{6}\text{Li}')$ data presented here provide an additional evaluation of the $E0$ strength on an odd-even nucleus in this region. By using a different probe and given the favorable background conditions for the $({}^{6}\text{Li},{}^{6}\text{Li}')$ reaction, it was possible to unambiguously settle the discussion about structure effects in the $A\approx90$ region.


\section{Experiment}

Inelastic scattering of ${}^{6}$Li particles  was measured at the Research Center for Nuclear Physics (RCNP), Osaka University. The present data are part of a  $({}^{6}\text{Li},{}^{6}\text{Li}'+\gamma)$ experiment to investigate the isovector spin-transfer response as a probe of the isovector magnetic dipole  transition strengths in the inelastic-scattering channel \cite{PhysRevC.98.015804}.  In that experiment, not only ${}^{6}\text{Li}'+\gamma$ events  but also singles ${}^{6}\text{Li}'$ events were recorded, the latter of which we report here. Note that the contribution from isovector excitations in the singles data is very small compared to the isoscalar excitations, as discussed in detail in Ref.~\cite{PhysRevC.98.015804}.
In the experiment,  a  ${}^{6}$Li beam was accelerated via the coupled operation of the azimuthally varying field (AVF) and ring cyclotrons to an energy of 100~MeV/u. The  beam was transported achromatically to the reaction targets with an energy spread of 1.5~MeV in full width at half maximum (FWHM). The beam intensity was monitored throughout the measurements and was approximately 1~pnA. The targets were self-supported  15.2-mg/cm$^2$ thick  ${}^{\text{nat}}$C  and 10.9-mg/cm$^2$ thick  ${}^{93}$Nb  foils. Inelastically scattered ${}^{6}$Li particles   were momentum analyzed and identified with the high-resolution magnetic spectrometer Grand Raiden \cite{FUJIWARA1999484}, which was placed at $0^\circ$ relative to the beam axis. The unreacted beam was stopped in a $0^\circ$ Faraday cup, which
was placed  at 12 m downstream of the focal plane.\par
The Grand Raiden focal-plane detectors consisted of two position-sensitive multiwire drift chambers (MWDCs) and three plastic scintillators \cite{PhysRevC.98.015804}. These detectors enabled the identification of the scattered particles as well as the reconstruction of their trajectories. The overall detection efficiency for ${}^{6}$Li particles was 74\%. By combining the positions in each MWDC, the angles in the dispersive and nondispersive directions were determined. A calibration measurement by using a sieve slit was used for the determination of the parameters of a ray-trace matrix for reconstructing scattering
angles at the target from position and angle measurements in the focal plane \cite{christhesis}. The ion optics of the spectrometer was tuned to run in the under-focus mode \cite{TAMII2009326} to optimize simultaneously the angular resolutions in the dispersive [2.8-mrad (FWHM)] and nondispersive [10.3-mrad (FWHM)] planes. The momentum reconstruction of the ${}^{6}$Li ejectiles was calibrated by measuring the elastic-scattering peak from the ${}^{93}\text{Nb}({}^{6}\text{Li},{}^{6}\text{Li})$ reaction at several magnetic rigidities.\par
The plastic scintillators in the focal plane (with thicknesses of 3, 10, and 10 mm) served to extract energy-loss signals and the time of flight (ToF) that was measured relative to the radio-frequency signal of the AVF cyclotron. A 12-mm aluminum plate was placed in between the second and the third scintillators in order to improve the particle-identification capabilities. ${}^{6}$Li particles were stopped in this plate, whereas $d$ and $\alpha$ particles from ${}^{6}$Li breakup punched through and deposited energy in the third scintillator. Therefore,  a veto signal from this detector was used  to remove the contribution from ${}^{6}$Li breakup in the offline analysis.\par
Inelastic scattering measurements  at angles  in the range of $0^\circ < \theta_{\text{lab.}}< 2^\circ$ were achieved. The magnetic-rigidity settings of the spectrometer covered excitation energies ($E_x$) from 10 to 40~MeV. Absolute cross sections were determined on the basis of calibration runs in which the beam intensity was measured with a Faraday cup inserted before the reaction target in between runs. The normalizations from these calibration data were then applied to the other runs. The uncertainty in the absolute cross sections determined with this procedure was estimated at 20\%, which was dominated by the read-out accuracy of the Faraday cup in the calibration runs due to the relatively low current.


\section{Data Analysis}
\begin{figure}[!ht]
\centering
\includegraphics[width=0.5\textwidth]{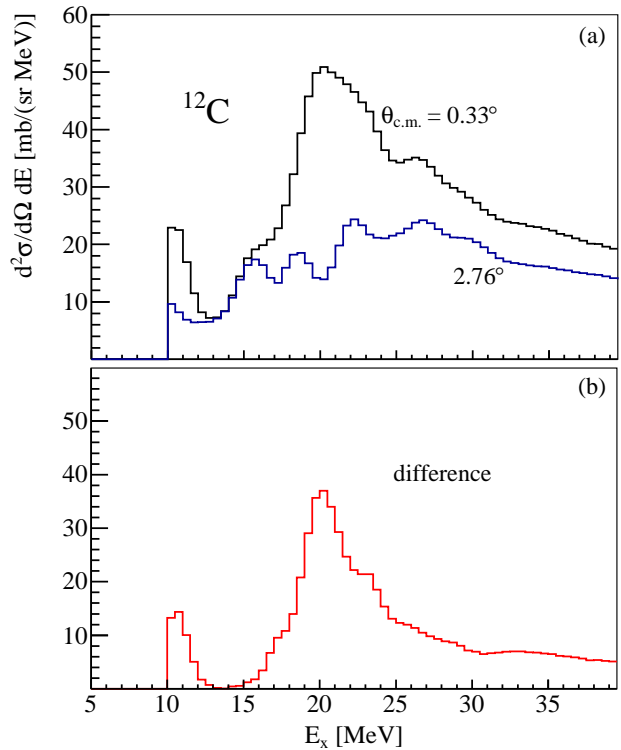}
\caption{\label{diff12c} (color online). Double-differential cross section  for the $({}^{6}\text{Li},{}^{6}\text{Li}')$ reaction on  ${}^{12}$C.  (a) Spectra for scattering angles at $0.33^\circ$ (close to the ISGMR maximum) and $2.76^\circ$ (near the first minimum of the ISGMR angular distribution).  (b) The difference between the above two spectra, which represents the ISGMR cross section  in ${}^{12}$C.  }
\end{figure}

\begin{figure}[!ht]
\centering
\includegraphics[width=0.5\textwidth]{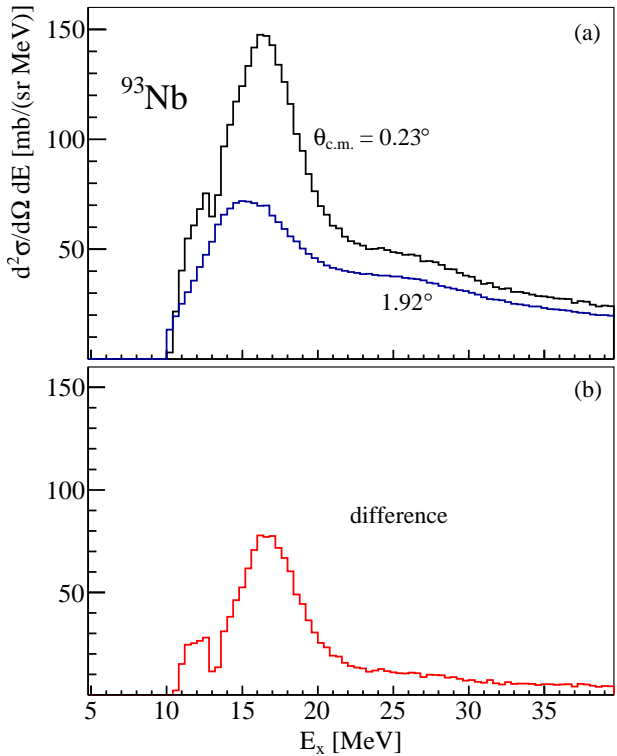}
\caption{\label{diff93nb} (color online). Double-differential cross section  for the $({}^{6}\text{Li},{}^{6}\text{Li}')$ reaction on  ${}^{93}$Nb.  (a) Spectra for scattering angles at $0.23^\circ$ (close to the ISGMR maximum) and $1.92^\circ$ (near the first minimum of the ISGMR angular distribution).  (b) The difference between the above two spectra, which represents the ISGMR cross section  in ${}^{93}$Nb. }
\end{figure}

Double-differential cross sections for inelastic scattering off ${}^{12}$C and ${}^{93}$Nb were binned into 0.5-MeV wide intervals in $E_x$. The angular acceptance was divided into $0.5^\circ$-wide bins covering scattering angles up to $2^\circ$ in the laboratory frame. Figs.~\ref{diff12c}(a) and \ref{diff93nb}(a) show examples of double-differential cross sections at different scattering angles for ${}^{12}$C and ${}^{93}$Nb, respectively. Note that inelastically scattered ${}^{6}$Li particles for $E_x<10$~MeV were not detected because of the magnetic rigidity settings of the spectrometer.  A rough estimate for the monopole contribution to the excitation energy spectra can be obtained by subtracting the spectrum around the angle where the first minimum of the ISGMR angular distribution is expected from the spectrum near $0^\circ$ scattering angle (where the ISGMR strength is maximal) \cite{BRANDENBURG198729}. As all the other multipolarities have relatively flat distributions in this angular region, the difference spectrum more or less represents  the ISGMR cross section (see Figs.~\ref{diff12c}(b) and \ref{diff93nb}(b)). Clearly, the monopole contributions to the measured excitation-energy spectra at forward scattering angles are very strong.

A more quantitative way to extract the giant resonance strengths is from a  multipole-decomposition analysis (MDA)  \cite{BONIN1984349}. The MDA  was performed for each bin in $E_{x}$ by fitting the differential cross section with a linear combination of distorted-wave Born approximation (DWBA)  distributions for angular momentum transfers of $\Delta L=0\textendash3$. These theoretical cross sections were obtained assuming 100\% exhaustion of the energy-weighted sum rule (EWSR) for each multipole. The DWBA calculations were performed with the code \texttt{CHUCK3} \cite{chuck3}. The  transition potentials were obtained using a double-folding formalism with  the  M3Y-Paris  nucleon-nucleon interaction \cite{KOBOS1984205}. A density-dependent term (BDM3Y1) was included to account for the reduction of the strength of the interaction as the density of the medium increases \cite{KHOA19956}. The ground-state density distribution used in the folding analysis for ${}^{12}$C was taken from Ref.~\cite{PhysRevC.33.40}, while for ${}^{93}$Nb it was taken  from Ref.~\cite{FRICKE1995177}. The depths of the resulting optical model potentials (OMP) were adjusted to fit the elastic scattering data from Ref.~\cite{refId0} (${}^{12}$C and ${}^{90}$Zr). The systematic uncertainty due to the choice of the OMP in the MDA was estimated to be 2\%. Parameterizations for  the transition densities, sum rules and deformation factors   employed in this analysis are described in Ref.~\cite{harakeh2001giant}.\par
As the isovector giant dipole resonance (IVGDR) can also be excited in inelastic  scattering of an isoscalar probe \cite{PhysRevLett.62.16,PhysRevLett.66.1287}, this small component of the cross section (below 10\% of the total at energies from 10 to 25~MeV) was subtracted from each angular distribution before performing the MDA. The IVGDR contribution was calculated on  the basis of the Goldhaber-Teller model \cite{harakeh2001giant} in conjunction with photonuclear cross-section data \cite{RevModPhys.47.713}.
The isovector spin-transfer response in the present data was studied by tagging the inelastic-scattering channel with the 3.56~MeV decay $\gamma$ ray, as discussed in Ref.~\cite{PhysRevC.98.015804}. However, it was shown that this contribution is small (less than 10\%) in comparison with the excitation of isoscalar giant resonances, and  does not significantly impact the present analysis.\par

\begin{figure*}[!ht]
\centering
\includegraphics[width=1.0\textwidth]{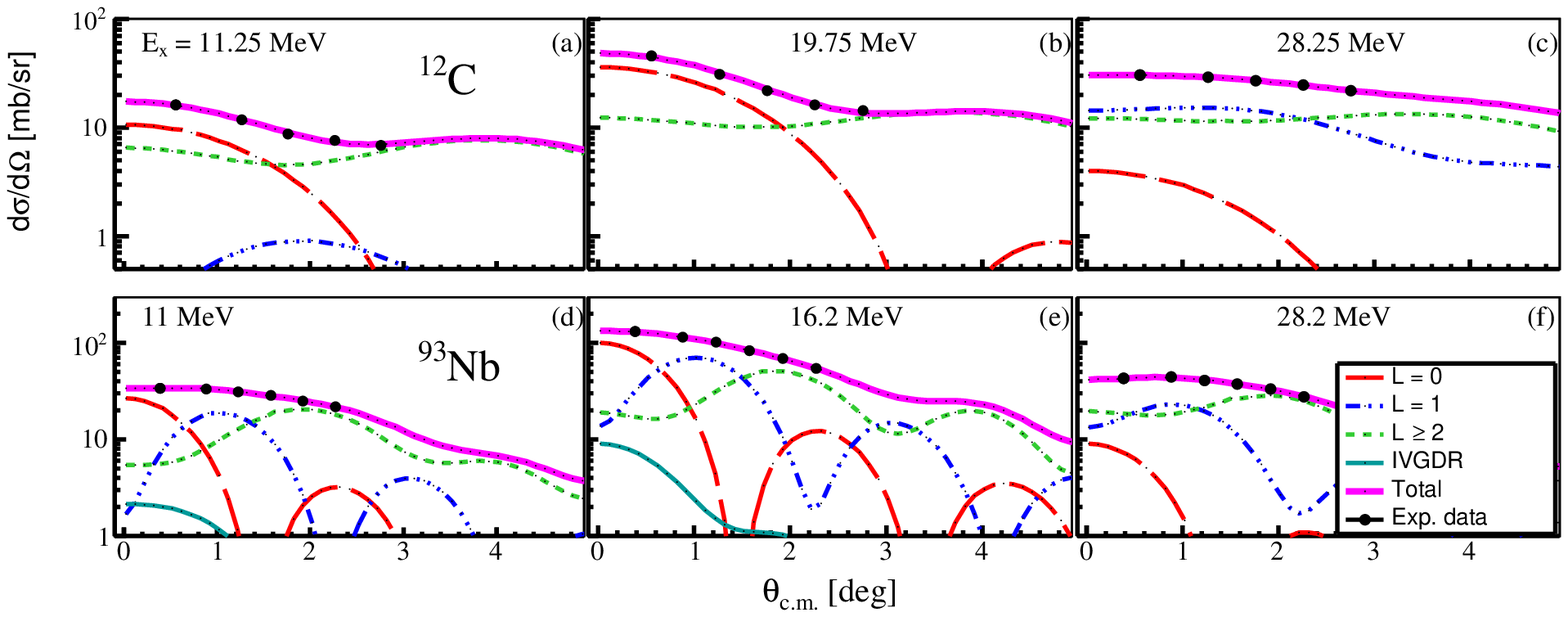}
\caption{\label{adts} (color online). Angular distributions for the $({}^{6}\text{Li},{}^{6}\text{Li}')$ reaction on  ${}^{12}$C (top) and ${}^{93}$Nb (bottom) at different excitation energies. The experimental data were fitted with MDA using DWBA calculations for angular-momentum transfers of $\Delta L=0\textendash3$. The contribution of the IVGDR in ${}^{93}$Nb was subtracted from the data before applying the MDA (see the text for details). }
\end{figure*}

\begin{figure*}[!ht]
\centering
\includegraphics[width=1.0\textwidth]{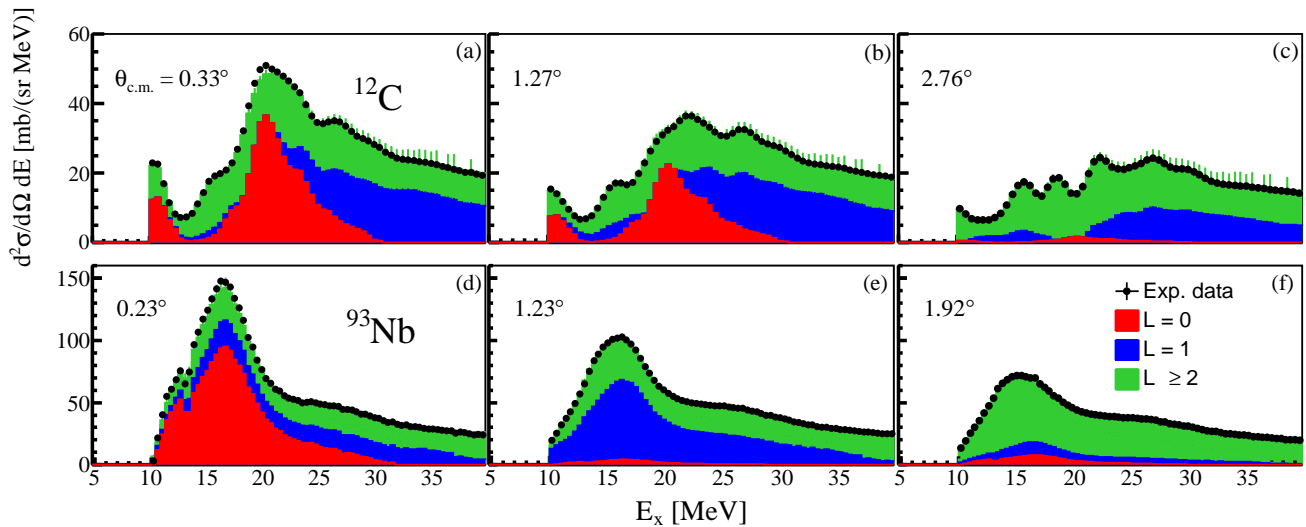}
\caption{\label{ddcs} (color online). Double-differential cross sections for ${}^{6}$Li scattering on ${}^{12}$C (top) and ${}^{93}$Nb (bottom) at different angles.  The colors of the stacked histograms represent the contributions from excitations with different angular-momentum transfers. }
\end{figure*}

Fig.~\ref{adts} shows the multipole components fitted to  angular distributions at different selected $E_x$ for ${}^{12}$C (top) and ${}^{93}$Nb (bottom). The double-differential cross sections for different angular bins as a function of $E_x$ are shown in Fig.~\ref{ddcs}. The stacked histograms represent the contributions of each multipolarity extracted from the MDA. As can be seen, the $L=0$ component for each nucleus is dominant at very forward angles. The monopole strength in the excitation-energy spectrum for ${}^{12}$C is concentrated in the range from 14 to 30~MeV. The cross section in this region has an  asymmetric distribution with a maximum around $E_x=20$~MeV, which is very similar  to Fig.~\ref{diff12c}(b). The monopole cross section for ${}^{93}$Nb  extends from 10 to 32~MeV with a weighted mean value at 17.5~MeV.

\section{Results and Discussion}

\subsection{${}^{12}$C}

In light nuclei, the $L=0$ strength distributions are usually fragmented over a wide energy range and only exhaust a small fraction of the EWSR \cite{PhysRevC.68.014305,PhysRevC.64.064308,GUPTA2015343,PhysRevC.65.034302}. As can be noticed from  Fig.~\ref{ddcs}, the ${}^{12}$C monopole cross section has a fragment below $E_x=14$~MeV, similar to that observed in $\alpha$-scattering experiments \cite{PhysRevC.68.014305, PhysRevC.84.054308}.  The multipole strength of ${}^{12}$C below 14~MeV was studied in Ref.~\cite{PhysRevC.84.054308}, where a $2^+$ state was found at $E_x\sim 10$~MeV submerged into a broad $0^+$ peak. This $2^+$ component  may be interpreted as the $2^+$ excitation of the Hoyle state and the $\alpha$-condensate state \cite{PhysRevC.84.054308}. Although in the present experiment only excitation energies above 10 MeV could be observed, the data are in excellent agreement with Ref.~\cite{PhysRevC.84.054308} since a combination of monopole and quadrupole contributions to the excitation-energy spectrum between 10 and 14 MeV were extracted. A fit with a Gaussian function was performed for the peak partially observed at $E_x\sim 10$~MeV. The fitted peak has a mean value at  $E_x=10.56$~MeV and  RMS  (root mean square) width of 1.03~MeV, which is in excellent agreement with the $0_4^+$ state observed in Ref.~\cite{PhysRevC.84.054308}.  Also, several $2^+$ states were observed in the $L=2$ cross section extracted from the MDA. Table~\ref{table1} lists the fitted mean $E_x$ and RMS width for each state. The same $2^+$ states were excited  in an $\alpha$-scattering experiment, as reported in Ref.~\cite{PhysRevC.68.014305}  (see Lit. $E_x$ column in Table~\ref{table1}).

\begin{table}[!ht]
\caption{\label{table1} Isoscalar $E0$ and $E2$ strengths extracted from the MDA. }
 \begin{ruledtabular}
\def\arraystretch{1.2}
\begin{tabular}{c c   c   c  c c}
Nucleus & Range &   Mean $E_x$  &  RMS Width & Lit. $E_x$& $L$  \\
 & [MeV] &   [MeV]  & [MeV] & [MeV] & [$\hbar$]   \\ \hline \hline
$^{12}$C &10.0\textendash 14.0  & 10.56\footnotemark[1] & 1.03 &10.56\footnotemark[3] & 0  \\
 &14.0\textendash 30.0  & 21.53\footnotemark[2] & 2.82 & & 0 \\
 &13.8\textendash 17.3  & 15.67\footnotemark[1] & 0.86 &15.42\footnotemark[4] & 2 \\
 &17.3\textendash 20.1  & 18.48\footnotemark[1] & 0.75 &18.90\footnotemark[4] & 2 \\
 &20.1\textendash 24.4  & 21.97\footnotemark[1] & 1.08 &22.31\footnotemark[4] & 2 \\
 &24.4\textendash 29.0  & 26.11\footnotemark[1] & 1.67 & & 2 \\
 &&&&&\\
$^{93}$Nb &10.0\textendash 32.0  & 17.51\footnotemark[2] & 4.12 & & 0  \\ 
\end{tabular}
 \end{ruledtabular}
 \footnotetext[1]{Gaussian fit}
 \footnotetext[2]{Weighted mean value}
 \footnotetext[3]{From Ref.~\cite{PhysRevC.84.054308}}
 \footnotetext[4]{From Ref.~\cite{PhysRevC.68.014305}}
\end{table}

\begin{figure}[!ht]
\centering
\includegraphics[width=0.5\textwidth]{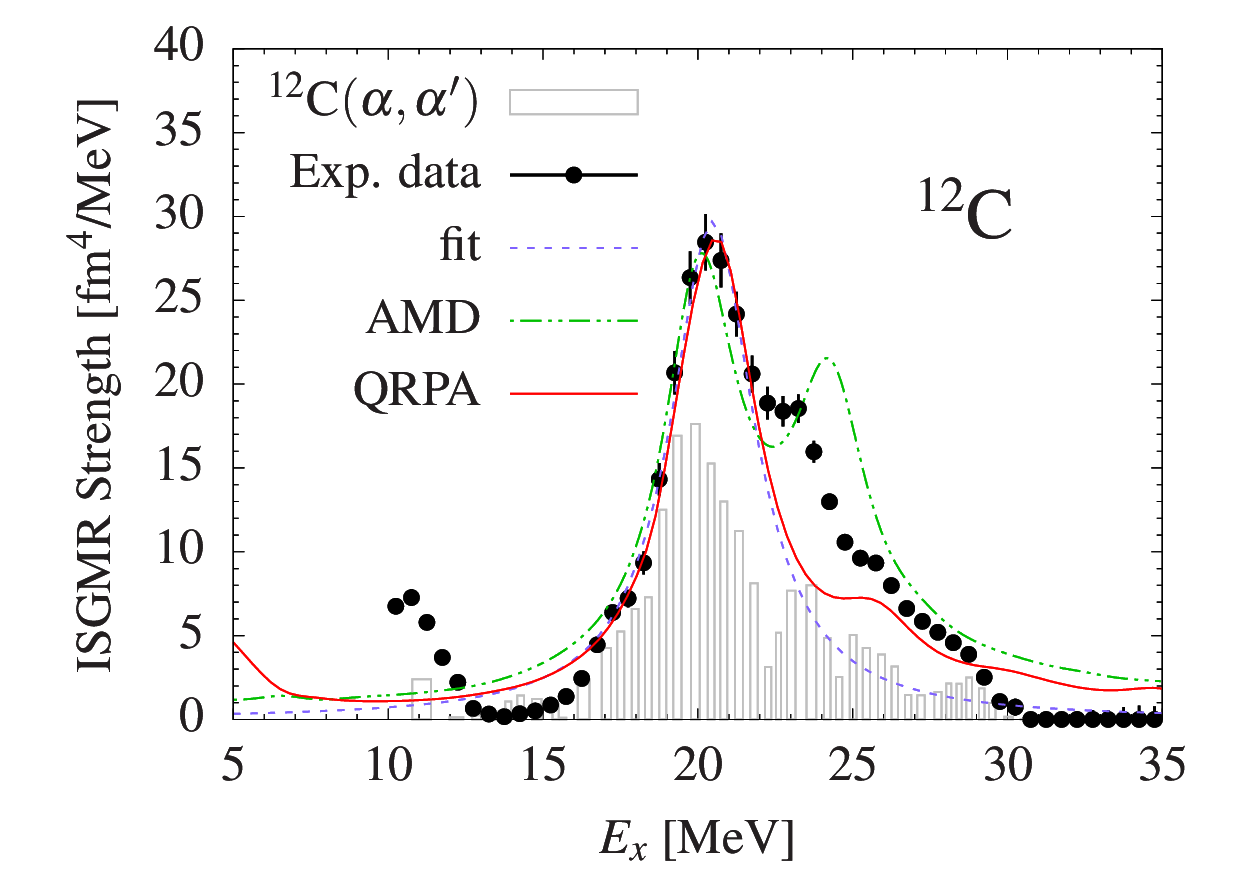}
\caption{\label{mono12c} (color online). ISGMR strength of ${}^{12}$C. The data are compared with results from a ${}^{12}\text{C}(\alpha,\alpha')$ experiment reported in Ref.~\cite{PhysRevC.68.014305}. The dashed line is a Lorentzian function fitted to the data  in the $E_x$ range  $14\textendash 22$~MeV. The dash-dotted line corresponds to a theoretical calculation based on antisymmetrized molecular dynamics extracted from Ref.~\cite{PhysRevC.82.034307}. The solid red line is  a QRPA calculation (shifted by a constant factor $\Delta = 1$~MeV) as described in the text. }
\end{figure}

In order to obtain information about the ISGMR it is necessary to extract the $S_0(E_x)$ strength distribution from the fitted $a_0(E_x)$ coefficients ($L=0$ component at $E_x$ extracted with MDA) as \cite{harakeh2001giant,GARG201855}

\begin{equation}
 S_0(E_x) = \frac{2\hbar^2 A \langle r^2 \rangle}{m  E_x}a_0(E_x),
\end{equation}
where $m$, $A$ and $\langle r^2 \rangle$ are the nucleon mass, mass number and the mean-square radius of the ground-state density, respectively. The extracted ISGMR strength distribution for ${}^{12}$C is presented in Fig.~\ref{mono12c}. This $E0$ distribution exhausts $52 \pm 3^\text{(stat.)} \pm 8 ^\text{(sys.)}$\% of the EWSR in the energy range $10\textendash30$~MeV, and $49 \pm 3^\text{(stat.)} \pm 8 ^\text{(sys.)}$\% from 14 to 30~MeV. The latter value is 22\% larger than the 27(5)\% reported in Ref.~\cite{PhysRevC.68.014305} from an $\alpha$-scattering  experiment. For comparison, the $(\alpha,\alpha')$ data from that experiment are also plotted in Fig.~\ref{mono12c}.

As can be seen, the shape of the previously extracted distribution is similar, but a large part of the ISGMR strength is missing above 18~MeV. Apparently, the model  employed in Ref.~\cite{PhysRevC.68.014305} to subtract the continuum+instrumental background strongly affected the extracted multipole strengths and removed contributions at high excitation energy. The maximum of the $E0$ distribution from $({}^{6}\text{Li},{}^{6}\text{Li}')$ data is located at $E_\text{m}=20.47(6)$~MeV, which was obtained from a Lorentzian fit over the $E_x$ range  $14\textendash 22$~MeV (see Table~\ref{table2}).  The centroid energy of the ISGMR, which represents the collective frequency of the compression mode,  is usually calculated from the ratio of different moments of the distribution. For example, the ISGMR centroid energy can be calculated as $\sqrt{m_1/m_{-1}}$ (in the hydrodynamical model) or $\sqrt{(m_3/m_1)}$ (in the generalized scaling model), where $m_k$ is the $k$-th moment of the distribution defined as $m_k=\sum_i E_i^k S_0(E_i)$ \cite{STRINGARI1982232}. The respective centroids, calculated in the energy range from 10 to 30~MeV, are presented in Table~\ref{table2}.

\begin{table}[!ht]
\caption{\label{table2} EWSR, Lorentzian fitted parameters and energy moments of the ISGMR strengths for ${}^{12}$C and ${}^{93}$Nb. Only statistical uncertainties are included.}
 \begin{ruledtabular}
\def\arraystretch{1.2}
\begin{tabular}{c c c c c c }
Nucleus & EWSR &  $E_\text{m}$  & $\Gamma$ & $\sqrt{m_1/m_{-1}}$ & $\sqrt{m_3/m_1}$   \\
 & [\%] &   [MeV]  & [MeV] & [MeV] & [MeV]   \\ \hline \hline
$^{12}$C & 52(3) & $20.47(6)$\footnotemark[1]   & $3.3(2)$  & $20.7(6)$  & $22.2(8)$  \\ 
$^{93}$Nb & 92(4) & $16.81(7)$\footnotemark[2]  & $7.1(2)$ & $17.5(1)$  & $19.9(1)$    \\ 
\end{tabular}
 \end{ruledtabular}
 \footnotetext[1]{fitted range: $14\textendash 22$~MeV}
 \footnotetext[2]{fitted range: $11\textendash 24$~MeV}
\end{table}

A  small bump in the ${}^{12}$C monopole distribution is observed at $E_x=23$~MeV. The location of this bump is also consistent with the $(\alpha,\alpha')$ data, as can be seen in Fig.~\ref{mono12c}. The existence of a second peak in the ISGMR around this energy has been predicted from antisymmetrized molecular dynamics (AMD) calculations in Ref.~\cite{PhysRevC.82.034307}. The calculation was scaled  and folded with a Lorentzian distribution (2~MeV width to account for the energy spread at high excitation energy) for a better comparison with the data (see Fig.~\ref{mono12c}). Note that the AMD calculation already has 1~MeV width that comes from a time filter applied to compensate for the finite integration in the Fourier transform \cite{PhysRevC.82.034307}. The qualitatively good agreement of this model with the data is an indication of an $\alpha$-cluster oscillation component at high excitation energies. In this theoretical approach, the ${}^{12}$C monopole resonance can be understood as a combination of a coherent vibration (breathing mode) and an oscillation of three $\alpha$ clusters around the center of the nucleus in a triangular structure. In the AMD, the $\alpha$-cluster vibration is observed as a common translation in the radial component of the single-particle wave functions with respect to their ground-state value. The breathing mode is obtained by the dynamical evolution of the widths of the single-particle wave functions (fermionic molecular dynamics).

The microscopic mean-field-based quasiparticle random-phase
approximation (QRPA) provides a good description of the collective states in nuclei \cite{ring2004nuclear}. A consistent axially-symmetric-deformed Hartree-Fock-Bogolyubov (HFB)+QRPA approach using the D1M Gogny interaction \cite{PhysRevLett.102.242501, sophie1}, has been employed to calculate the ISGMR strength of ${}^{12}$C.  Here, the single-particle wave functions are expanded on an optimized harmonic oscillator basis. In this approach, the intrinsic deformation of ${}^{12}$C ground state ($\beta=-0.4$) was predicted by the HFB calculations as the minimum of the potential energy surface. The resulting model space configuration allowed to build coherent two-quasiparticle (2-$qp$) excitations and the respective transition probabilities in the QRPA calculation.  The QRPA energies were shifted by a constant factor $\Delta = 1$~MeV to account for a small energy displacement originated in the coupling between $qp$ states and phonons. This effect  has been systematically observed  in comparisons to the giant dipole resonance peak with D1M+QRPA calculations for a wide range of nuclei \cite{PhysRevC.94.014304,PhysRevC.98.014327}. The resulting  QRPA monopole distribution was folded with a Lorentzian function of 3~MeV width, as shown in Fig.~\ref{mono12c} (solid line). As can be seen, the calculation is fairly consistent with the experimental data. The deformation effects lead to a double-peak distribution concentrated in energies from 15 to 30~MeV.  In particular, the calculation successfully describes the data around the centroid energy and also the asymmetric tail at high excitation energies.

\subsection{${}^{93}$Nb}

\begin{figure}[!ht]
\centering
\includegraphics[width=0.5\textwidth]{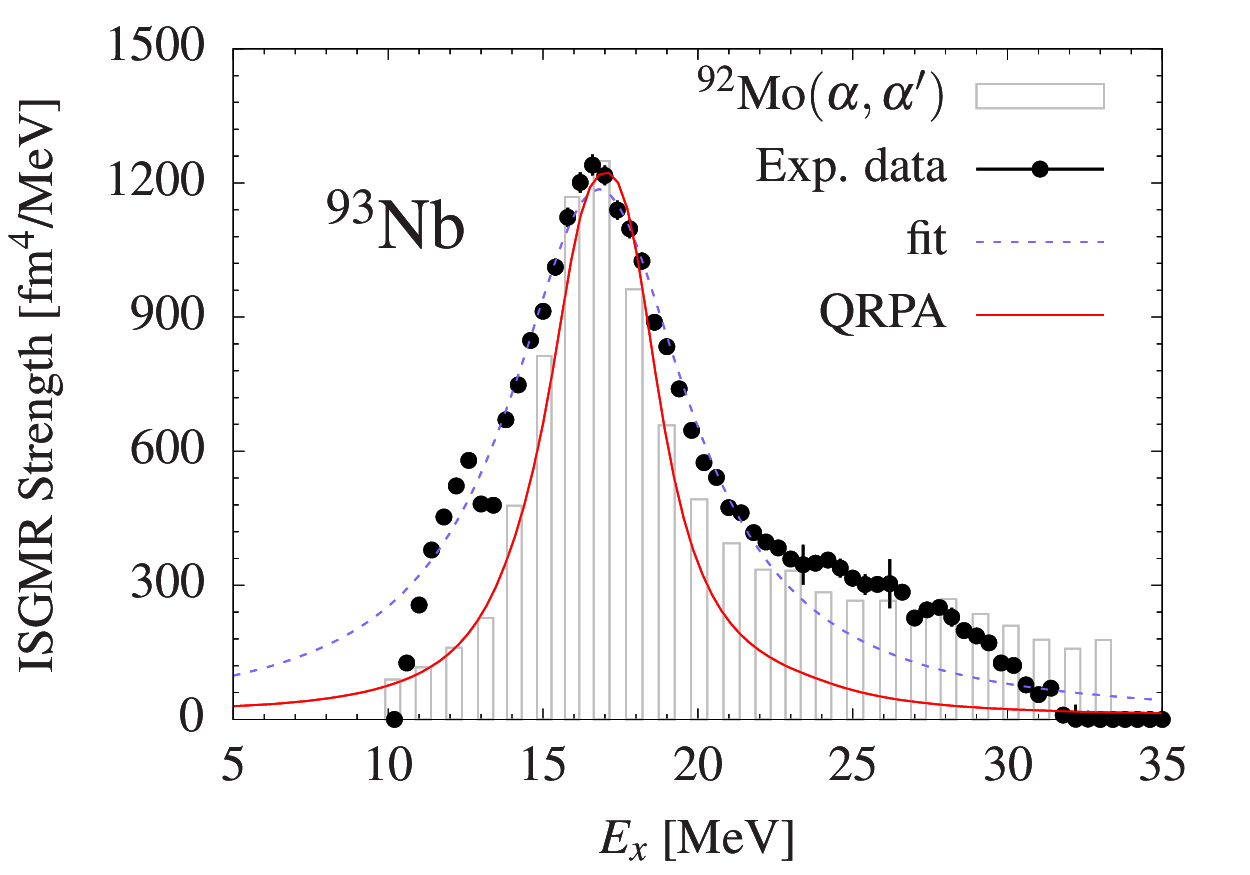}
\caption{\label{mono93nb} (color online). ISGMR strength of ${}^{93}$Nb. The data are compared with results from a ${}^{92}\text{Mo}(\alpha,\alpha')$ experiment reported in Ref.~\cite{GUPTA2016482}. The dashed line is a Lorentzian function fitted to the data  in the $E_x$ range  $11\textendash 24$~MeV. The solid red line corresponds to a QRPA calculation (shifted by a constant factor $\Delta = 1$~MeV) as described in the text. }
\end{figure}

The ISGMR strength extracted for ${}^{93}$Nb is presented in Fig.~\ref{mono93nb}. In contrast with ${}^{12}$C, the  ISGMR distribution of ${}^{93}$Nb is not fragmented. Almost  all  $E0$ strength of ${}^{93}$Nb is concentrated around 17~MeV, exhausting  $92 \pm 4^\text{(stat.)} \pm 10 ^\text{(sys.)}$\% of the EWSR in the energy range $10\textendash32$~MeV. Recently, ISGMR studies in the $A\approx90$ region  suggest a significant dependence of the centroid energy on nuclear structure in these nuclei \cite{PhysRevC.88.021301}. However, the results from a different experiment \cite{GUPTA2016482} indicate  that the fluctuations of the ISGMR centroids for nuclei in this region are too small to invoke  a shell-structure effect.  The $E0$ strength of ${}^{92}$Mo extracted from  $\alpha$-scattering in Ref.~\cite{GUPTA2016482} is also plotted in Fig.~\ref{mono93nb} for a better comparison with our results. As can be seen, the present data are quite consistent with this ${}^{92}$Mo ISGMR distribution. The ${}^{93}$Nb distribution at low energy (between 10 to 14~MeV) exhibits a larger strength (about 5\% more of the EWSR). It has been observed that in deformed nuclei the ISGMR strength separates into two components because of the coupling to the isoscalar giant quadrupole resonance \cite{PhysRevC.60.067302,harakeh2001giant,PhysRevC.68.064602}. Thus, the small $E0$ component  at low energy ($\sim 13$~MeV) in ${}^{93}$Nb can possibly be associated with   deformation effects on the ISGMR distribution, similar to the recent results for the neutron-rich  ${}^{94,96}$Mo nuclei \cite{Howard2019}. Nevertheless,   the deformation effects  are  negligible for a comparison with the ISGMR centroid energies of $A\approx90$ nuclei, as seen in  Fig.~\ref{mono93nb} and also suggested in Ref.~\cite{Howard2019}.

A Lorentzian fit over the $E_x$ range $11\textendash 24$~MeV gives a centroid located at 16.81(7)~MeV (see Table~\ref{table2}), which is in good agreement with the results for the even-even ${}^{90,92}$Zr and ${}^{92,94,96}$Mo nuclei from Refs.~\cite{GUPTA2016482,Howard2019}. The nuclear incompressibility  extracted from the $\sqrt{(m_3/m_1)}$ centroid in the scaling model for ${}^{93}$Nb corresponds to $K_A=178(2)$~MeV, and it is also consistent within the error bars with the values reported in Refs.~\cite{GUPTA2016482,Howard2019}. Therefore, the present results confirm that the nuclear structure does not have a significant impact on the ISGMR strength distribution in the $A\approx90$ region, even for the odd-even ${}^{93}$Nb nucleus. 

The ISGMR strength of ${}^{93}$Nb was also calculated by using a fully consistent axially-symmetric-deformed HFB+QRPA approach with the D1M Gogny interaction. The HFB ground state was obtained with the blocking technique for $K^\pi=9/2^+$, and an oblate deformation ($\beta=-0.02$)  that minimized the potential energy. The calculation was performed on a harmonic oscillator basis that includes 13 major shells to reduce by 1~MeV the energy shift  between the experimental and theoretical results \cite{PhysRevC.94.014304}. Coupling  between monopole and quadrupole states was enabled in order to  account for splitting effects in the $E0$ strength.  Similar to the adopted procedure for the QRPA calculation of ${}^{12}$C, the energies were shifted by a constant factor of $\Delta = 1$~MeV and the strength was folded with a Lorentzian function of 3-MeV width to compare with the experimental data. The  QRPA result for the ISGMR strength  of ${}^{93}$Nb  is presented in  Fig.~\ref{mono93nb}. As can be seen, the calculation is consistent with the experimental data. As usual, the QRPA reproduces well the centroid energy and total strength of the giant resonance but not the width. To reproduce the total width of the ISGMR, it would be necessary to enlarge the configuration space by adding  4-$qp$ excitations \cite{KAMERDZHIEV20041,PhysRevC.79.034309}.


\section{Summary}

Inelastic scattering of ${}^{6}$Li particles at 100~MeV/u off ${}^{12}$C and ${}^{93}$Nb  have been measured at scattering angles between $0^\circ$ and $2^\circ$.
Measurements free of instrumental background and the  very favorable resonance-to-continuum ratio of ${}^{6}$Li scattering  enabled the precise extraction of the   ISGMR distribution in ${}^{12}$C and ${}^{93}$Nb. A multipole-decomposition analysis was performed in the excitation-energy range from 10 to 35~MeV to extract the  contributions from transitions associated with the transfer of different units of angular momentum transfer.  The isoscalar $E0$ strength was strongly excited in the angular range covered in the present experiment. The  ISGMR distribution obtained from the ${}^{12}$C data exhausts 22\% more of the EWSR than previous  measurements with $\alpha$ scattering have reported. The difference is likely due to the method previously employed to remove background from the data. The present data are qualitatively well described by  an AMD calculation that takes into account  vibration modes from  three $\alpha$ clusters,  and also by an axially-symmetric-deformed HFB+QRPA calculation.\\
The extracted ${}^{93}$Nb ISGMR strength is concentrated at $E_x=17$~MeV and exhausts $92 \pm 4^\text{(stat.)} \pm 10 ^\text{(sys.)}$\% of the EWSR in the energy range $10\textendash32$~MeV. About 4\% of the strength is located in a small bump at 13~MeV, which can be associated with a deformation of the ISGMR. The ${}^{93}$Nb ISGMR distribution was compared with other results from $A\approx90$ nuclei. The centroid energies of the $E0$ distributions from these nuclei are consistent, and   no nuclear-structure effects were observed for the ISGMR energy location. A large-scale deformed QRPA calculation was performed to obtain the  ${}^{93}$Nb monopole strength. This theoretical calculation is consistent with the experimental data.



\section*{Acknowledgements}
We thank the staff of RCNP for their tireless efforts in preparing the CAGRA array, the Grand Raiden spectrometer, and the ${}^{6}$Li beam. J.C.Z.  thanks the support by Fundaç\~ao de Amparo a Pesquisa do Estado de S\~ao Paulo (FAPESP) under Grant No.~2018/04965-4. This material was based on work supported by the National Science Foundation under Grants No.~PHY-1430152 (JINA Center for the Evolution of the Elements), No.~PHY-1565546, No.~PHY-1913554 and No.~PHY-1713857, by the US DOE under Contract No.~DE-AC02-06CH113567, by German BMBF grant No.~05P19RDFN1, by the International Joint Research Promotion Program of Osaka University, by the DFG under
Contract No.~SFB~1245, and by the Hirose International Scholarship Foundation.

\bibliography{bibliography}  

\end{document}